\documentclass[proceedings, preprint]{rmaa}

\usepackage{paralist}
\usepackage{lipsum}

\usepackage{psfrag,color}
\usepackage{graphicx,amsfonts,fancyhdr}
\usepackage[dvipsnames]{xcolor}

\SetYear{2015}
\SetConfTitle{RevMexAA (Serie de Conferencias), 00, 1–4}

\title{The SONG prototype: Efficiency of a robotic telescope} 

\author{
  M. F. Andersen,\altaffilmark{1} 
  F. Grundahl,\altaffilmark{1}
  A. H. Beck,\altaffilmark{1}
  and P. Pall\'e\altaffilmark{2}}

\altaffiltext{1}{Stellar Astrophysics Centre, Aarhus University (Denmark)}
\altaffiltext{2}{Instituto de Astrof\'{i}sica de Canarias (IAC), Tenerife (Spain)}

\shortauthor{Andersen, Grundahl, Beck \& Pall\'e}
\shorttitle{SONG: Efficiency}

\listofauthors{W. J. Henney, A. Collaborator, \& L. Author}
\indexauthor{Henney, W. J.}
\indexauthor{Collaborator, A.}
\indexauthor{Author, L.}

\abstract{The Stellar Observations Network Group prototype telescope at the Teide Observatory has been operating in scientific mode since March 2014. The first year of observations has entirely been carried out using the high resolution echelle spectrograph. Several asteroseismic targets were selected for scientific and technical verification. A few bright subgiants and one red giant were chosen since the oscillations in these stars have large amplitudes and the periods long enough to easily be detected. These targets would also be used for evaluation of the instruments since long term observations of single targets would reveal potential problems. In this paper the performance of the first robotic SONG node is described to illustrate the efficiency and possibilities in having a robotic telescope.}

 \resumen{El telescopio prototipo del Grupo de la Red de Observaciones Estelares (SONG) ha operado en modo cient\`ifico desde marzo de 2014. El primer a\~no de observaciones se ha dedicado por completo al espectr\`ografo. Varios objetos de inter\`es astros\`ismico se han observado para verificaci\`on cient\`ifica y t\`ecnica. Algunas estrellas subgigantes brillantes y una gigante roja fueron elegidas para prueba ya que las oscilaciones en estas estrellas tienen grandes amplitudes y los periodos son lo suficientemente largos para ser detectados. Estos objetos ser\`an usados para evaluar los instrumentos ya que las observaciones a largo plazo de los objetos de estudio podr\`ian presentar algunos problemas. En este art\`iculo describimos como opera el primero de los telescopios de la Red SONG para ilustrar la eficiencia y las capacidades de un telescopio robótico.}

\addkeyword{Robotic telescope}
\addkeyword{Spectroscopy}
\addkeyword{Bright stars}
\addkeyword{Radial velocities}

\begin{document}
\maketitle

\section{Introduction}
\label{sec:intro}

Astronomical observatories are entering an era where on-site man power is less and less required. The technological advancements during the last decade have made robotic observatories feasible and most importantly affordable. With high speed internet connections and off-the-shelf network controlled hardware components robotic telescopes appear more frequently and old-fashioned observatories controlled by an operator at night are getting less common. \\
\\
The prototype telescope in the SONG project, Stellar Observations Network Group, was inaugurated on the 25$^{th}$ October 2014. This telescope is the first in a network of small 1\,m telescopes which will be placed strategically on Earth to make 24 hours coverage of single stars possible \citep{concept}. The prototype is located at the Teide Observatory (OT) in the Iza\~na mountain on Tenerife.\\
\\
Since March 2014 the telescope has been operating in scientific mode collecting high quality spectroscopic data using the high resolution echelle spectrograph.\\
\\
The complete SONG observatory is fully robotic and can be remotely controlled from anywhere through the Internet if needed. The location at OT is perfect for robotic observatories with highly skilled on-site maintenance staff and a dedicated high speed internet connection to the main land of Spain. Together with a working power grid these components are combined to form a perfect location for a robotic telescope. The sky quality at OT is one of the best in the world which means that this location is not only perfect because of the infrastructure but also because of the potential to deliver the best possible data. The median seeing per year at Tenerife is below 1$''$ \citep{seeing}.

\section{Telescope performance}
\label{sec:telescope}

\begin{figure}[!h]
\begin{center}
  \includegraphics[width=0.8\columnwidth]{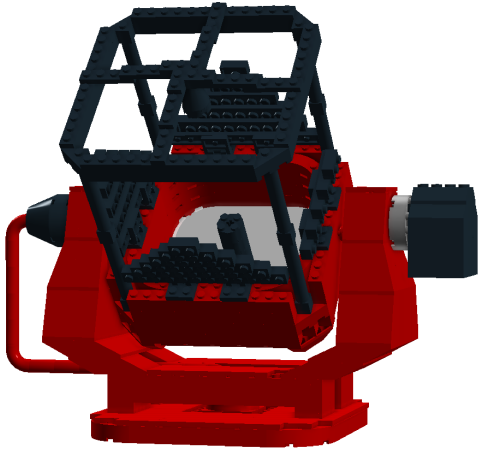}
  \caption{LEGO model of the SONG prototype telescope. The pipe on the left is the coud\'e train and the the black box on the right is the lucky imaging unit.}   
  \label{fig:lego} 
\end{center}
\end{figure}

The 1\,m SONG telescope manufactured by ASTELCO Systems GmbH is an Alt-Az Cassegrain-Nasmyth mounted fast response telescope with a slewing speed of up to 20 $^{\circ}/s$. With a pointing model including about 100 measurements a root mean square (RMS) of 3-4$''$ can be achieved which is a necessity for SONG which has a small field of view of approximately 30 arcseconds. In Fig.~\ref{fig:lego} a model of the prototype telescope can be seen.

\noindent The telescope is operated through the open Telescope Software Interface (TSI) delivered by ASTELCO and makes simultaneous connections possible. This means it is possible to perform several actions at the same time which if used can minimize the overhead when operating the telescope.\\
\\
The 5\,cm thin primary mirror is corrected using active optics (AO). A lookup table has been created using a Shack-Hartmann wave-front sensor to make measurements at different altitudes. When observing the AO is updated at every repoint and checked every minute and updated if the lookup table dictates it. \\
\\
The telescope has a moveable M3 which makes it possible to use both Nasmyth focii for instruments. On one side we have the entry to the coud\'e train which guides the light to the spectrograph placed in the accompanying shipping container. In the other Nasmyth port the lucky imaging (LI) unit is located \citep{li}. Switching between the to instruments by rotating M3 180$^{\circ}$ takes less than one minute.

\section{Spectrograph performance}
\label{sec:spectrograph}

\begin{figure}[!h]

\begin{center}
  \includegraphics[width=1.0\columnwidth]{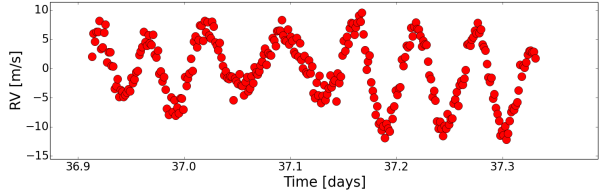}
  \caption{One full night with approximately 10 hours of radial-velocity measurements of $\gamma$ Cephei. The periodic variations are due to stochastically exited solar-like oscillations.}
  \label{fig:gammaceph}
\end{center}
\end{figure}

The high resolution echelle spectrograph has been fully operational since its installation in 2012.
In Table~\ref{tab:slit} the characteristics of the different possible slits are displayed.\\
\\
The spectrograph performs as expected and produces high quality spectra with the Iodine or Thorium-Argon (ThAr) calibration method.
A few examples on the radial-velocity precision of the SONG spectrograph using the Iodine method is displayed in Table~\ref{tab:spec}. The table shows the short term precision. The best velocity precision is reached for cool, slowly rotating bright stars.
The long term precision is of the order of 5\,m/s determined by using radial-velocity standard stars.\\
\\
In Fig.~\ref{fig:gammaceph} radial-velocity measurements of $\gamma$~Cephei are shown. The observations of this full night show a velocity precision of around 1\,m/s which is excellent for asteroseismic investigations. 

\begin{table}[!t]\centering
  \setlength{\tabnotewidth}{\columnwidth}
  \tablecols{4}
  \setlength{\tabcolsep}{2.0\tabcolsep}
  \caption{SONG spectrograph slits} \label{tab:slit}
  \begin{tabular}{cccc}
    \toprule
    \# & \multicolumn{1}{c}{Width [$\mu$m]} & \multicolumn{1}{c}{Width [$''$]}  & \multicolumn{1}{c}{Resolution\tabnotemark{a}}\\
    \midrule
    1 & \o20, \o100 & - & -\\
    2 & \o20 & \o0.69 & -\\  
    4 & 100 & 3.44 & 40,000\\   
    5 & 60 & 2.06 & 60,000\\  
    6 & 45 & 1.55 & 80,000\\  
    7 & 36 & 1.24 & 90,000\\  
    8 & 30 & 1.03 & 100,000\\  
    9 & 25 & 0.86 & 112,000\\  
    \bottomrule
    \tabnotetext{a}{The resolution is the rounded median of the entire spectrum with lowest resolution in the blue and highest in the red.}
  \end{tabular}
\end{table}

\begin{table}[!t]\centering
  \setlength{\tabnotewidth}{\columnwidth}
  \tablecols{3}
  \setlength{\tabcolsep}{2.8\tabcolsep}
  \caption{SONG spectrograph performance sample} \label{tab:spec}
  \begin{tabular}{lcc}
    \toprule
    Target & \multicolumn{1}{c}{Magnitude} & \multicolumn{1}{c}{Precision} \\
    \midrule
    Procyon & 0.34 &  3 m/s\\
    $\gamma$ Cephei    & 3.22 & 1 - 2 m/s\\
    $\mu$ Herculis    & 3.42 & 3 m/s\\
    46LMi    & 3.83 & 1.5 - 2 m/s\\
    HD109358    & 4.25 & 5 m/s\\
    HD185144 & 4.67 & 3 m/s\\    
    \bottomrule
    \tabnotetext{}{All targets were observed with a 36 micron wide slit (1.24$''$) which gives a resolution of approximately 90,000 and using the Iodine method}
  \end{tabular}
\end{table}

\section{Robotic observations}
\label{sec:robotic}

\begin{figure*}[!t]
\begin{center}
  \includegraphics[width=0.8\textwidth]{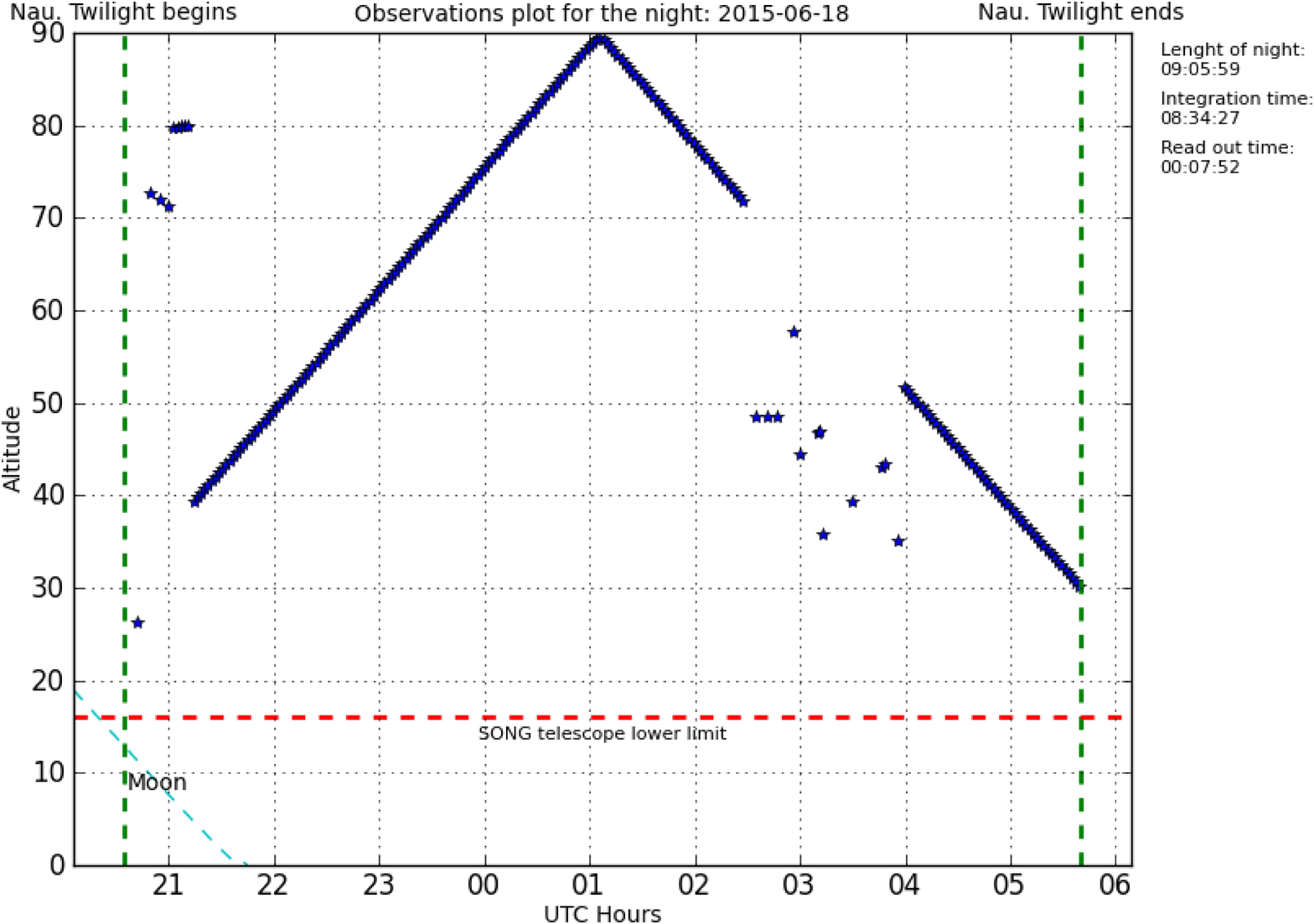}
  \caption{A summary plot of the observed data during one full summer night. Each asterisk corresponds to an acquired spectrum. This night 9 different targets were observed and the primary asteroseismic target was observed twice.}
\label{fig:summary} 
\end{center}
\end{figure*}

For each observing night a list of targets is entered into a central database which is then replicated to the site at Tenerife and it will be at all other sites \citep{mfa}. The targets are entered with an observing window and a priority. Before submission a check function is constructed to evaluate the list and a plot shows a theoretical estimate of the following night of observations. If the user is satisfied the list can be submitted and the rest will be handled automatically.\\
\\
After each night an automatically generated e-mail will be sent to the technical staff with a summary of the observations that given night. An attached plot showing the details are illustrated in Fig.~\ref{fig:summary}. The figure shows a very typical summer night with SONG; One primary asteroseismic target, a few radial velocity standard stars and some filler programs.

\begin{figure}[!t]
\begin{center}
  \includegraphics[width=0.8\columnwidth]{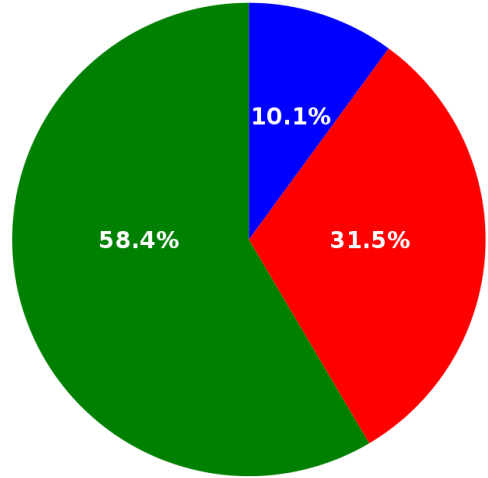}
  \caption{The complete duty cycle of the SONG prototype telescope during the first year of operation from March 2014 and a year onwards. \textbf{Green}: Observed percentage of possible hours, \textbf{Red}: Downtime due to weather, \textbf{Blue}: Technical downtime.}
    \label{fig:duty1}
\end{center}
\end{figure}

\noindent The asterisks mark the altitude at which each spectrum is acquired as a function of time during the night. The green dashed lines indicates the beginning and ending times at which we start and stop observing. This is when the Sun is at -6 degrees below the horizon which right after/before sunset/sunrise defines the beginning/end of nautical twilight. The integration time where light is collected on the CCD detector is displayed together with the total possible observing time and the total read out time of the CCD. This illustrates the efficiency of the telescope and the robotic software. We see the total overhead that specific night was about 30 minutes where almost 8 minutes was used on reading out the CCD. This gives about 22 minutes of overhead where we changed targets 10 times. Two targets were observed with a ThAr sandwich and the rest using Iodine. The overhead per ThAr reference frame is about one minute which leaves less than two minutes of overhead per target. These two minutes include slewing, acquisition of the target and moving motors for the specific observing mode for each target.

\section{Duty cycle the first year}
\label{sec:duty}

From March 2014 until the first proposal run on April 2015 the SONG telescope at the Teide Observatory was used for verification of the scientific and technical specifications. \\
\\
The total percentage of time used for observations the first year compared to the total hours of usable night time was 58.4\,\%. 31.5\,\% of the total time the telescope was closed down due to bad weather and the last 10.1\,\% was down time because of technical improvements and issues. This is illustrated in Fig.~\ref{fig:duty1}.
The biggest contributions to the technical time during the first year of operation were; in July a server failure in the middle of the holiday period which caused 7 days of down time, in August 2014 the dome failed due to ungreased and broken rollers for the azimuth rotation which stopped operations for 10 days and in February 2015 the lucky imaging system was moved from one Nasmyth to the other and because of rewiring, alignment tests, etc., which caused a week of down time. Without these events the technical time would be about 3\,\% which includes several nights for testing the lucky imaging system, several half nights for creating the AO lookup table by ASTELCO and one full night because during Christmas eve the OT is closed.

\begin{figure}[!t]
\begin{center}
  \includegraphics[width=1.0\columnwidth]{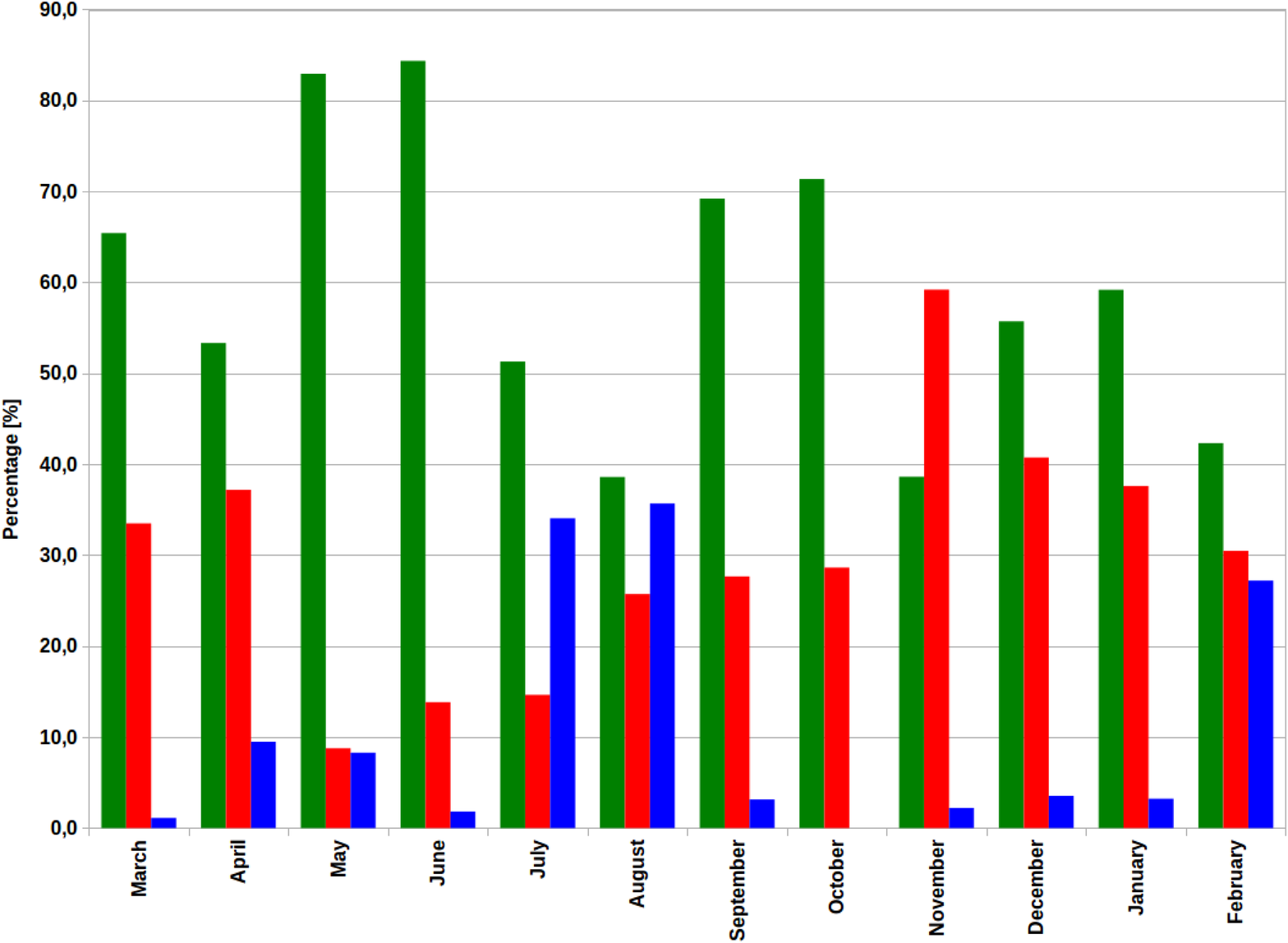}
  \caption{The monthly duty cycle of the SONG prototype telescope during the first year of scientific operation. \textbf{Green}: Observed percentage of possible hours, \textbf{Red}: Downtime due to weather, \textbf{Blue}: Technical downtime.}
  \label{fig:duty2} 
\end{center}
\end{figure}

\section{Results}
\label{sec:results}

\begin{figure}[!h]
\begin{center}
  \includegraphics[width=1.0\columnwidth]{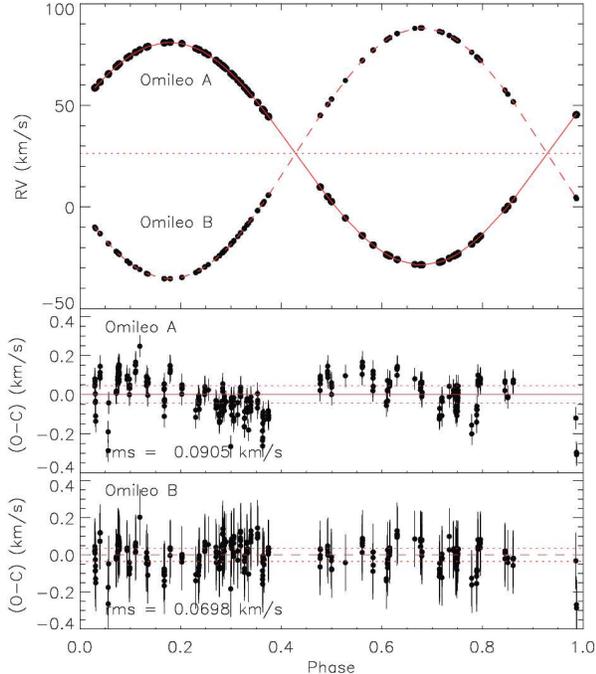}
  \caption{Top: Radial-velocity phase curve for Omicron Leonis obtained during
2014 and early 2015. Bottom-top: Residuals after subtracting the keplerian orbit (red) from the primary component. Bottom-bottom: Residuals after subtracting the keplerian ordit (red) from the secondary component.}
  \label{fig:omileo}
\end{center}
\end{figure}

During the first year of scientific operation several asteroseismic targets and many filler programs were observed. One of the filler programs was the spectroscopic binary star system Omicron Leonis. In Fig.~\ref{fig:omileo} the radial-velocity phase curve is shown. This target was observed with the ThAr method and the radial-velocities shown were determined using broadening functions.\\

\noindent A total of 333 spectra have been analysed \citep{beck}. The Keplerian orbit and residuals are shown. Note that
the overall scatter in the two single-component solutions is
90 and 70\,m/s for the primary and secondary components, respectively.\\
\\
Although hard to see each epoch consists of (typically) 5 consecutive
exposures (analysed separately). A ThAr spectrum was obtained before
and after each sequence of 5 exposures. Within each sequence the radial-velocity
scatter is 33 and 46\,m/s for the primary and secondary components, respectively. We are
currently working to improve the data-reduction pipeline and calibration
methods to reach this level also over long time-scales.  In the final orbital
solution the minimum masses are determined with a precision of 0.001 M$_{\bigodot}$.

\section{Future}
\label{sec:future}
\begin{figure}[!h]
\begin{center}
  \includegraphics[width=1.0\columnwidth]{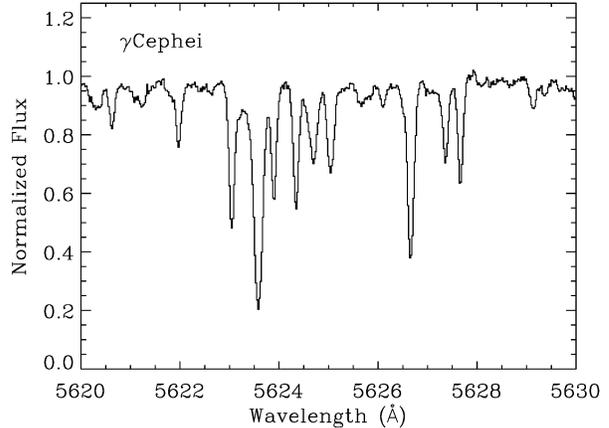}
  \caption{Part of a spectrum acquired with the Chinese SONG node. The target $\gamma$ Cephei was selected since good comparison spectra from the prototype at OT has already been obtained.}
  \label{fig:china}
\end{center}
\end{figure}

The second SONG node at the Delingha Observatory in China is in the process of hardware tweaking. The first light using the telescope together with the spectrograph was obtained in autumn 2015. Analysis of the first acquired spectrum showed that the specifications were met and the resolution is equal to the one from the SONG spectrograph placed on the prototype node at Tenerife.
A small part of an acquired spectrum using the Chinese SONG node is shown in Fig.~\ref{fig:china}. The plot shows the normalized flux in a narrow wavelength range to illustrate the quality at which the SONG spectrograph in China can deliver.\\ 
 
\noindent Implementation of the robotic software at the Chinese node is still to come and extensive complete system tests will be needed.\\

\section*{Acknowledgements}
We would like to acknowledge the Villum Foundation and the Carlsberg Foundation for the support on building the SONG prototype on Tenerife. The Stellar Astrophysics Centre is funded by The Danish National Research Foundation (Grant DNRF106) and research is supported by the ASTERISK project (ASTERoseismic Investigations with SONG and Kepler) funded by the European Research Council (Grant agreement n. 267864). We also gratefully acknowledge the support by the Spanish Ministry of Economy Competitiveness (MINECO) grant AYA2010-17803.\\

\end{document}